\documentclass{ws-procs975x65}

\newcommand{\be}{\begin{equation}}
\newcommand{\ee}{\end{equation}}
\newcommand{\beqa}{\begin{eqnarray}}
\newcommand{\eeqa}{\end{eqnarray}}
\newcommand{\bea}{\begin{eqnarray}}
\newcommand{\eea}{\end{eqnarray}}
\newcommand{\rom}[1]{\mathrm{#1}}
\newcommand{\sr}{{\textsf{R}}}
\newcommand{\sm}{{\textsf{M}}}
\newcommand{\sj}{{\textsf{J}}}
\newcommand{\sa}{{\textsf{A}_\textsf{H}}}
\newcommand{\lp}{\left(}
\newcommand{\rp}{\right)}
\newcommand{\eg}{{\it e.g.,}\ }
\newcommand{\mc}[1]{\mathcal{#1}}
\newcommand{\al}{\alpha}

\begin{document}

\begin{flushright}
AEI-2010-037
\end{flushright}

\title{ \Large  ON THE BLACK HOLE SPECIES\\
(BY MEANS OF NATURAL SELECTION)}

\author{ MARIA J. RODRIGUEZ,}

\address{Max-Planck-Institut f\"ur Gravitationsphysik,\\
Albert-Einstein-Institut, 14476 Golm, Germany\\
$^*$E-mail: maria.rodriguez@aei.mpg.de}

\begin{abstract}
Recently our understanding of black holes in D-spacetime dimensions, as
solutions of the Einstein equation, has advanced greatly. Besides the well
established spherical black hole we have now explicitly found other species of
topologies of the event horizons. Whether in asymptotically flat, AntideSitter
or deSitter spaces, the different species are really non-unique when $D\geq5$.
An example of this are the black rings. Another issue in higher dimensions that
is not fully understood is the struggle for existence of regular black hole
solutions. However, we managed to observe a selection rule for regular solutions
of thin black rings: they have to be balanced i.e. in vacuum, a neutral
asymptotically flat black ring incorporates a balance between the centrifugal
repulsion and the tension. The equilibrium condition seems to be equivalent to
the condition to guarantee regularity on the geometry of the black ring
solution. We will review the tree of species of black holes and present new
results on exotic black holes with charges.
\end{abstract}


\bodymatter

\section{Introduction}\label{intro}

\indent Black holes are the most elementary and intriguing objects of General Relativity (GR). The fact that the effects of the spacetime curvature are dramatic in their presence explains why it is relevant studying these systems.

String/M-theory is currently the best candidate for a unified theory of all interactions and, in particular, is expected to describe quantum gravity. One of the most surprising outcomes of the theory was its prediction of the dimensionality of spacetime. This, perhaps contrary to expectation, was required to be ten rather than four. As its low-energy limit, higher dimensional GR can be regarded as a powerful tool to gain insights into the more fundamental theory, as well as deserving further study in its own right. As it has been argued, higher dimensional black holes could form at very high energies and, actually, form at the Large Hadron Collider (LHC) \cite{ArkaniHamed:1998nn} at CERN. Bearing in mind the deep reasons to consider GR in dimensions higher than four, we aim to analyze its most remarkable solutions, black holes, in a higher-dimensional setting.

The vast number of black holes\footnote{Note, incidentally, that the name {\em black hole} was coined by John Archibald Wheeler in 1967 \cite{Wheeler}.} that exist in the Universe, usually lying at the centre of galaxies, are exactly represented by the black hole solution found by Roy Kerr \cite{Kerr}, a neutral(electrically uncharged)\footnote{If electric charges are allowed, the only possible four-dimensional black hole solutions are the so-called Reissner-Nordstrom (non-rotating) and Kerr-Newman black holes (rotating) \cite{Mazur:1982db,BuntingTesis}.} rotating Schwarzschild black hole \cite{Schwarzschild} in four spacetime dimensions. This species is unique \cite{Israel,Carter} as well as the topology of the event horizon which can only be spherical \cite{Hawking1}, namely $S^2$ and characterized only by its mass and angular momentum (called \emph{charges} in this context).

These objects have a theoretical counterpart in higher spacetime dimensions. But, unlike in four, in higher dimensions there are other examples of black hole solutions with new exciting properties. In fact, the four-dimensional uniqueness theorems break down for $D > 4$ and, accordingly, horizon topologies other than spheres can, and do indeed, arise. Almost 100 years after the discovery of the first black hole solution, the catalogue of different species (exact solutions) of black holes shows a very rich structure but seems far from being complete -- in flat space, besides the Myers and Perry (MP)black hole \cite{Myers:1986un} in five dimensions, there are also the black ring \cite{Emparan:2001wn,Pomeransky:2006bd}, the black saturn \cite{Elvang:2007rd}, the di-ring \cite{Iguchi:2007is,Evslin:2007fv} and the bicycling black ring \cite{Elvang:2007hs}. The list is as well enlarged by axisymmetric black holes known approximately such as the higher dimensional black rings \cite{Emparan:2007wm} and its more general cousins the blackfolds \cite{Emparan:2009cs}.

But before plunging into the different examples of black objects let us go back to the theory we will be interested in, namely GR in $D$ dimensions. The central field equation of GR in vacuum is the commonly called Einstein equation
\begin{eqnarray}\label{EinsteinCosmo}
R_{\mu\nu}-\frac{1}{2}g_{\mu\nu}\;R+\frac{(D-2)}{2}\,\Lambda\;g_{\mu\nu}=0\,
\end{eqnarray}
($\mu , \nu =1,2,...,D$), of a remarkable simplicity which nevertheless hides an extraordinary mathematical complexity. The {\it geometry} of spacetime is encoded in the metric $g_{\mu \nu}$, which features explicitly and within the Ricci tensor $R_{\mu \nu}$ and scalar $R$, that measure the curvature of spacetime. We will allow, in general, for a cosmological constant $\Lambda$. Typically, $\Lambda=0, \pm (D-1)L^{-2}$, where $L$ is the radius of the curved space.
It follows from (\ref{EinsteinCosmo}) that vacuum solutions are either Ricci-flat ($R_{\mu \nu} =0$), if $\Lambda =0$, or Einstein ($R_{\mu \nu} =\Lambda g_{\mu \nu}$), otherwise. Immediate solutions include $D$-dimensional Minkowski space (if $\Lambda =0$), $D$-dimensional de Sitter space, dS (if $\Lambda>0$), and $D$-dimensional Anti-de Sitter space, AdS (if $\Lambda<0$). Depending on the value of the cosmological constant, the black hole solutions of (\ref{EinsteinCosmo}) that we will consider will typically asymptote to one of these three spaces, either in all $D$ directions or in a smaller number of directions (that we will call transverse). At the practical level, this will translate in the imposition of appropriate boundary conditions.

The difficulty in solving Einstein's equations (\ref{EinsteinCosmo}) increases as the number $D$ of spacetime dimensions does too. Indeed, the larger number of degrees of freedom, $\frac{1}{2}(D-2)(D-1)-1$, carried by the unknown metric $g_{\mu \nu}$ to be solved for makes of (\ref{EinsteinCosmo}) an increasingly involved system of coupled, nonlinear, partial differential equations. On the other hand, in a higher-dimensional spacetime \emph{there is more room} than in four dimensions for solutions to be able to display richer features. For example, solutions can now rotate in up to $N=[\frac{D-1}{2}]$ independent rotation planes, the number of Casimir operators (independent angular momenta $J^2_i$) of the spacelike rotation group $SO(D-1)$. Despite the complexity of the problem, as we have already mentioned, many black hole solutions are known and therefore it is natural to seek a classification.

There are many ways to perform a classification of black holes.
They can be classified according to their boundary conditions and charges (mass and angular momentum) or, instead, in terms of the topology of the event horizon. This is, perhaps, a more interesting classification since it is the event horizon what exclusively distinguishes the black holes from other stellar objects, that lack it and that consequently display  more conventional properties. As we have just mentioned, in four spacetime dimensions the topology of a black hole's event horizon is restricted to be the sphere $S^2$. Interestingly enough, this restriction drops if spacetime is allowed to have a higher number of spacelike directions, in which case  much richer possibilities do indeed arise.
We will present a catalogue according to the topology of their event horizon and point out a selection rule that might explain the struggle for existence of regular black hole solutions.

From here on we will be mainly concerned with stationary (time independent) black hole solutions of higher-dimensional GR. In the following section, \ref{sec:topological}, the possible event horizon topologies of higher-dimensional single black holes are reviewed. The explicit metrics of the known examples are recorded in section \ref{sec:blacksolutions} and the general properties of multi black holes are discussed in section \ref{sec:multi}. In section \ref{sec:curvedblacksolutions} we compile the explicit known solutions of black holes in curved backgrounds as well as the properties of more exotic cases such as the higher dimensional black ring. In the final sections we comment on the phase diagram, a selection rule for regular black hole solutions and discuss some open problems in the subject. This review is also intended as a brief guide to the higher-dimensional black hole solutions 

\vspace{2mm}\noindent \textbf{\it Conventions}

We will refer to ``extra" dimensions of spacetime when considering more than $4$ spacetime dimensions and we label them $n$ while setting $D=4+n$ where $D$ is the total number of spacetime dimensions. $G$ is Newton's constant in $D$ dimensions and the conventions used for unities are the speed of light, and the Planck and Boltzmann constants respectively $c=\hbar=k_B=1$. In order to make comparisons between the different asymptotically flat $D$-black holes we introduce dimensionless quantities (factoring out the mass $M$) for the spin $j$, the area $a_H$, the angular velocity $\omega_H$ and the temperature $\mathfrak{t}_H$ as
\begin{small}
\beqa
\label{jaHdef}
j^{n+1}=c_j\,
\frac{J^{n+1}}{GM^{n+2}} \,,\qquad
a_H^{n+1}=c_a\,\frac{\mathcal{A}^{n+1}}{(GM)^{n+2}}
~,\\
\omega_H =c_\omega \, \Omega_H
(GM)^{\frac{1}{n+1}} \,,\qquad
\mathfrak{t}_H = c_{\mathfrak{t}}\, (GM)^{\frac{1}{n+1}}\, T_H\,,
\eeqa
\end{small}
where the numerical constants (defining $\Omega_n$ the $n$-volume of a unit sphere) are
\begin{subequations}
\begin{small}
\be
c_j =\frac{\Omega_{n+1}}{2^{n+5}}\frac{(n+2)^{n+2}}{(n+1)^{\frac{n+1}{2}}}\,,\,\,\,
c_a=\frac{\Omega_{n+1}}{2(16\pi)^{n+1}}(n+2)^{n+2}
\left(\frac{n}{n+1}\right)^{\frac{n+1}{2}}\,,
\ee
\be
c_\omega=\sqrt{n+1}\left(\frac{n+2}{16}\Omega_{n+1}\right)^{-\frac{1}{n+1}}
\,,\,\,\,
c_{\mathfrak{t}} =  4\pi \sqrt{ \frac{n+1}{n}}
\left( \frac{n+2}{2} \Omega_{n+1}\right)^{-\frac{1}{n+1}}
\,.
\ee
\end{small}
\end{subequations}
We find convenient to introduce different dimensionless magnitudes for asymptotically (A)dS black holes, denoted by sans-serif fonts,
corresponding to quantities measured in units of the cosmological radius
$L$ or `cosmological mass' scale $L^{D-3}/G$. For instance, for the
$S^1$-radius, mass, angular momentum and horizon area of the ring we
define
\begin{small}
\be\label{dless}
\sr=\frac{R}{L}\,,\qquad \sm =\frac{GM}{L^{D-3}}\,,\qquad
\sj=\frac{GJ}{L^{D-2}}\,,\qquad \sa =\frac{A_H}{L^{D-2}}\,.
\ee
\end{small}
Equivalently, we might have set $L=1=G$, but the meaning of some formulas is
clearer if we retain $L$.

\vspace{-2mm}\section{Topological classification of black holes} \label{sec:topological}

Black objects in any dimension can be characterized and classified according to the topology of their event horizon. On the one hand, the classification of neutral, asymptotically flat, \emph{static} black holes (non-rotating solutions with null Killing vector fields on the horizon) is simple and complete. The Schwarzschild-Tangherlini black hole has been proved \cite{Israel,Carter,Bunting,Gibbons:2002bh} to be the only allowed static black hole in all dimensions $D \geq 4$, and the existence of static black holes with non spherical $S^{D-2}$ topologies is accordingly ruled out \footnote{The proof can be generalized to Einstein-Maxwell-dilaton theories \cite{Gibbons:2002bh}.}.

\begin{figure}
\centering
\vspace{-5mm}\includegraphics[height=6cm,width=6cm]{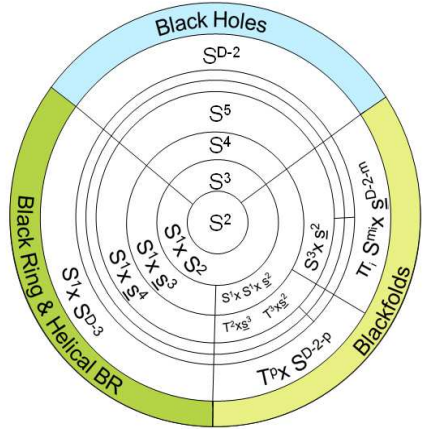}
\vspace{-2mm}
\caption{\small Summary of neutral stationary uni horizon black hole solutions in higher dimensions $D\geq4$ classified by its horizon topologies. In the center the four dimensional spherical Kerr black hole. Each of the outer shells represents one higher dimension i.e. in $D=5$, the first outer shell from the center, there are $S^3$ and $S^1 \times S^2$ topologies of the event horizons corresponding to the Myers-Perry black hole and the black ring (BR) and helical BR. For the higher dimensional black rings and blackfolds the solutions have been found perturbatively using the matched asymptotic expansions. These are characterized by two scales $R$ and $r_0$ corresponding to the radii of the spheres that conform its event horizon -- a large $S^{m_i}$ and a smaller $\underline{s}^{D-2-m}$ respectively. Note that for blackfolds $p\ge2$ and $\sum_i m_i =m$ while $\forall i\,  m_i$ are odd. The perturbative analytical metrics of black rings (with $S^1\times \underline{s}^{D-3}$ event horizons topologies) and p-tuboids (topologically $\mathbb{T}^p\times \underline{s}^{D-2-p}$) have been found in the \textit{thin} approximation, when $r_0<<R$. However, helical black rings and the more general blackfolds which are products of odd spheres are only known to linear order. The same summary is devised for black holes with asymptotical (A)dS boundary conditions.}\label{fig:negzones}\vspace{-5mm}
\end{figure}

In contrast, \emph{stationary} black holes (those with intrinsic rotation), can give rise to event horizons with more sophisticated topologies. The current status of the classification of stationary black holes by horizon topologies is far from being complete, and most of the higher-dimensional black hole solutions allowed in principle remain unknown. Let us first review what the situation is for stationary, asymptotically flat black holes in all dimensions (see figure \ref{fig:negzones} for a quick summary).
\begin{list}{\labelitemi}{\leftmargin=1.5em}
\item $D \leq 4$. There are no asymptotically flat black holes below four dimensions: the lowest dimensional known black hole is the well known Kerr black hole in four dimensions. As we discussed in section \ref{sec:blacksolutions}, this spherical \cite{Hawking1} ($S^2$) rotating black hole is the only one in $D=4$, and is uniquely characterized by its conserved charges \cite{Israel,Carter}.

\item $D=5$ ($n=1$). In five dimensions the situation is different since there are no equivalent general uniqueness theorems \cite{Morisawa:2004tc,Hollands:2007aj}. Now the possible topologies of the event horizon are not only $S^3$, but also $S^1 \times S^2$. The Myers-Perry black hole solution (the extension of the Kerr rotating black hole to higher dimensions) corresponds to the former, $S^3$, case. The latter possibility, $S^1 \times S^2$, is also realized in the black ring of Emparan and Reall \cite{Emparan:2001wn}, with one angular momentum, and that of Pomeransky and Sen'kov \cite{Pomeransky:2006bd}, with two angular momenta in orthogonal planes.

     These are, in fact, the only possibilities (aside from the Lens $L(p,q)$ topology where no explicit solution is known), according to rigidity theorems \cite{Galloway:2005mf,Galloway:2006ws}. All higher-dimensional black holes, regardless of their asymptotics, have been argued to be axisymmetric, that is, to display an axial $U(1)$ symmetry \cite{ Hollands:2006rj}. The existence of black hole solutions with exactly one $U(1)$ symmetry were conjectured in \cite{Reall:2002bh}. The first evidence for such a solution was provided in \cite{Emparan:2009cs} and dubbed helical black rings.

\item $D=6$ ($n=2$). When going one dimension further up, to six dimensions, the territory becomes vast and not many solutions are known. From cobordism theories \cite{Donaldson}, restrictions in the type of allowed topologies leave the open possibilities just to the following: $S^4$, $S^1\times S^3$ and $S^2\times\Sigma_g$ where $\Sigma_g$ is a genus $g$ Riemann surface \cite{Helfgott:2005jn} (for example the $S^2$, with $g=0$).
    For several years, the only known black hole solution in $D=6$ was, again, the MP black hole with two rotational symmetries, with an $S^4$ event horizon. The more extravagant topologies $S^1\times S^3$ were recently shown to be realized in the $D=6$, the \emph{thin black rings} \cite{Emparan:2007wm} and the \emph{thin helical black rings} (black rings in the weak gravity approximation, for which self gravitational effects are absent. See details in section \ref{sec:blacksolutions}). Also there is evidence of existence of black 2-tuboids with $\mathbb{T}^2\times S^2$ horizons topologies, representing a particular type of blackfold in six dimensions. Explicit $D=6$ black hole metrics realizing the remaining possibility, $S^2\times S^2$ for the event horizon, are unknown.

\item $D>6$ ($n>2$). In this case there are essentially no restrictions on the possible topologies. In fact, there are no analog rigidity theorems to restrict the topologies of the black holes' event horizons in  dimensions greater than 6. However, we do know of possible topologies: those realized by some explicitly known solutions. These include $S^{D-2}$ (realized in the Myers-Perry solutions), $S^1\times S^{D-3}$ (realized in the approximate solutions of thin black rings \cite{Emparan:2007wm}, see section \ref{sec:blacksolutions}), and finally, $\mathbb{T}^p \times S^{D-2-p} $ with $p\geq 2$ (realized in the black p-tuboids). There is also evidence of existence of black holes with  $ \prod_i S^{m_i}  \times S^{D-2-m}$ horizon topologies where $2\leq m_i\leq n$, where $  m_i\in \mathbb{N}_{odd}$ and $m=\sum_i m_i\leq n$. Collectively black holes with horizons that are products of spheres and tori (particularly dubbed black p-tuboids) will be called blackfolds here. Note that even-ball blackfolds are claimed to describe the ultraspinning MP black holes.
\end{list}
\vspace{-2mm}New evidence for the existence of exotic event horizon topologies (such as $S^2\times S^2$) can be found in \cite{Kleihaus:2009wh,Dias:2009iu,Dias:2010eu}.

Much less is known about black hole solutions which asymptotically approach global Anti-de Sitter space, AdS, at spatial infinity, the so called \emph{AdS black holes}. The reason is to be put down to the extra term that arises from the non-vanishing cosmological constant in the Einstein equation, which further complicates the problem. In fact, only AdS black holes with spherical horizons $S^{D-2}$ for $n\geq0$ extra dimensions have been known for a long time. These  include both the \emph{static} Schwarzschild-AdS solution \cite{Kottler,Witten:1998zw} and the \emph{stationary} Kerr-AdS black holes \cite{Hawking:1998kw,Gibbons:2004uw,Gibbons:2004js}. This situation has now changed with the discovery of the thin AdS black rings in all dimensions greater than four \cite{Caldarelli:2008pz}, the details of which are presented in section \ref{sec:curvedblacksolutions}.  Note that by topological censorship a four dimensional AdS/dS black ring can be ruled out.\vspace{-3mm}

\vspace{-1mm}\section{The asymptotically flat black hole solutions} \label{sec:blacksolutions}
This section includes the known metrics of the most general exact black objects with asymptotically flat boundary conditions. We review the asymptotically flat metrics of Myers-Perry black holes in all dimensions and of the doubly-spinning black ring in $D=5$. Then we proceed to recall the conserved charges for the approximate higher dimensional black rings, helical black rings and blackfolds. The comparison among them is performed in section \ref{sec:phase}.

\vspace{-3mm}\subsection{Black Hole}

The Myers-Perry black hole, the higher-dimensional counterpart of Kerr's black hole, exhibits rotation in all possible $N=[(D-1)/2]$ planes. In $D$ dimensions, its line element is given by \cite{Myers:1986un}
\beqa\label{MyersPerrySolution}
ds^2&=&- dt^2+ dr^2+r^2 d\alpha^2 \epsilon +\frac{\Pi F}{\Pi-\mu r}dr^2 +\sum_i (r^2+a_i^2)(d\mu_i^2+\mu_i^2 d\phi_i^2)\nonumber\\
&&+ \frac{\mu r}{\Pi\, F}\sum_i (dt+a_i\mu_i^2 d\phi_i)^2
\eeqa
where $i=1, \ldots , N$, $\epsilon=\mod_2 \,\,(D-1)$, $\mu_i$ are the direction cosines, $\phi_i$ the azimuthal angles, $\mu$ and $a_i$ are free parameters. The coordinates are restricted as $\sum_i\,\mu_i^2+\alpha^2 \epsilon=1$, and
$F=1-\frac{a_i^2 \mu_i^2}{r^2+a_i^2}$ and $\Pi=\prod_{i=1}^{N} (r^2+a_i^2)$.
There exists an event horizon, with spherical topology $S^{D-2}$, situated at $r_0$ the largest root of
\be\label{mueq}
\Pi-\mu \,r^{2-\epsilon}=0.
\ee
The black hole is characterized by the mass parameter $\mu$  and the rotation parameter $a_i$ by which we can express the thermodynamics
\begin{subequations}
\be\label{thermoMP}
M=\frac{\Omega_{D-2}}{16\pi G}(D-2) \,\mu ,\qquad S=\frac{\Omega_{D-2}}{4G} \,\mu \,r_0\,,
\ee
\be\label{TMPJOm} T_H=\frac{1}{2\pi r_0}\left(\,r_0^2\sum^{N}_{i=1}\frac{1}{r_0^2+a_i^2}-\frac{1}{1+\epsilon}\right)\,,
\ee
\be
J_i=\frac{\Omega_{D-2}}{16\pi G}a_i\,\mu,\qquad\Omega_i=\frac{a_i}{r_0^2+a_i^2}.
\ee
\end{subequations}
The event horizons of black holes are not at all rigid. On the contrary, they have been observed to be very elastic \cite{Emparan:2003sy}.
For large enough angular momenta the behaviour of some black
holes changes to that of extended black branes (black rings also exhibit a similar behavior for large spins and act like black strings -- see the following section and \cite{art} for details.)
Qualitatively, as the spin becomes large, the event horizon spreads out in the
plane of rotation and becomes a higher dimensional `pancake' approaching
the geometry of a black brane.
Our focus will be on the particular case in which the black hole has one large angular momenta and all others are zero. However, a detailed analysis of the more general situations in which black holes present black membrane phases can be found in \cite{art}.

An important simplification occurs in the ultra-spinning regime
of $J\to\infty$ with fixed $M$, which corresponds to
$a \rightarrow \infty$. Then (\ref{mueq}) becomes
\begin{equation}
\label{uslimit}
\mu \rightarrow a^2 r_0^{n-1}
\end{equation}
leading to simple expressions for the eqs.~(\ref{TMPJOm}) in terms of $r_0$ and
$a$, which in this regime play roles analogous to those of $r_0$ and $R$ for the
black ring. Specifically, $a$ is a measure of the size of the horizon along the
rotation plane and $r_0$ a measure of the size transverse
to this plane \cite{Emparan:2003sy}. In fact, in this limit
\be
\label{TMP2}
 M \to \frac{ (n+2)  \Omega_{n+2}}{16 \pi G}\; a^2 r_0^{n-1}
\,,\,\, S \to \frac{\Omega_{n+2}}{4 \pi G}\;a^2 r_0^{n} \,,\,\,
T_H \to \frac{n-1}{4 \pi r_0}
\end{equation}
take the same form as the expressions characterizing a black membrane extended
along an area $\sim a^2$ with horizon radius $r_0$.
This identification\footnote{The entropy corresponds precisely to a membrane of
planar area $\frac{\Omega_{n+2}}{\Omega_n}a^2$. This value also
gives the precise membrane mass once the dimension dependence of the
mass normalization is taken into account.} lies at the core of
the ideas in \cite{Emparan:2003sy}.
The reader may rightly wonder what happens to
\begin{equation}\label{JOm2}
J \to \frac{ \Omega_{n+2}}{ 8 \pi G}\;a^3 r_0^{n-1}
\,,\qquad \Omega_H \to \frac{1 }{a}\,,
\end{equation}
under this identification. Both turn out to disappear, since the
black membrane limit is approached in the region near the axis of rotation of
the horizon and so in the limit the membrane is static.
Observe that the value of (\ref{uslimit}), (\ref{TMP2}) and (\ref{JOm2}), are valid
up to $O(r_0^2/a^2)$ corrections.

The transition to this membrane-like regime is signaled by a qualitative
change in the thermodynamics of the MP
black holes.
At
\begin{small}
\be\label{onset}
\left(\frac{a}{r_0}\right)_{\rm mem}=\sqrt{\frac{D-3}{D-5}}\,,\vspace{-2mm}
\ee
\end{small}
the temperature reaches a minimum and
$\left(\partial^2 S/\partial J^2\right)_M$
changes sign. This point should not be considered
as a sign for an instability or a new branch but rather a transition to an infinitesimally nearby solution along the same family of solutions. The numerical evidence of \cite{Dias:2009iu} supports this connection with the zero-mode perturbation of the solution. For $a/r_0$ smaller than this value, the thermodynamic quantities
of the MP black holes such as $T$ and $S$ behave similarly to those of the Kerr
solution and so we should not expect any membrane-like behaviour.
However, past this point they rapidly approach the membrane results and develop a Gregory-Laflamme type of instability.

\vspace{-2mm}\subsection{Black Rings and helical black ring}

Black Rings, whose horizon exhibits an $S^1\times S^2$ topology, were first found by Emparan and Reall \cite{Emparan:2001wn}(see \cite{Emparan:2006mm} for a review).  Following this development, but now using the \emph{inverse scattering method} \cite{Belinsky:1971nt,Belinsky:1979,Belinski:2001ph}, Pomeransky and Sen'kov \cite{Pomeransky:2006bd} managed to build what is usually called the \emph{doubly spinning black ring} that had long been anticipated. It is balanced by angular momentum in the plane of the ring, with angular momentum also in the orthogonal plane corresponding to rotation of the two-sphere. The latter solution can be written as
\beqa\label{PomeranskiSolution}
ds^2&=&-\frac{H(y,x)}{H(x,y)}(dt+\Omega)^2+\frac{F(x,y)}{H(y,x)}d\phi^2+2\frac{J(x,y)}{H(y,x)}d\phi d\psi \nonumber\\
&&- \frac{F(y,x)}{H(y,x)}d\psi^2-\frac{2k^2 H(x,y)}{(x-y)^2(1-\nu)^2}\left(\frac{dx^2}{G(x)}-\frac{dy^2}{G(y)}\right)
\eeqa
Here, $k$, $\nu$, $\lambda$ are  parameters, $k_0=\nu(1-\lambda^2-\nu^2)$,  $k_1=\lambda(1-\lambda^2-3\nu^2+2\nu^3)$, the one-form $\Omega$ is defined as
$ \Omega=-\frac{2 k \lambda \sqrt{(1+\nu )^2-\lambda ^2}}{H(y,x)}( (1-x^2) y \sqrt{\nu}d\psi+
 \frac{(1+y)}{(1-\lambda +\nu)}(1+\lambda -\nu +x^2 y \nu (1-\lambda-\nu)+2\nu x(1-y))\,d\phi)$ and the functions $G$, $H$, $J$, $F$ as
 \beqa
 G(x)&=&(1-x^2)(1+\lambda x+\nu x^2)\,,\nonumber\\
 H(x,y)&=&1+\lambda ^2-\nu^2+2\lambda\nu (1-x^2)y+2x\lambda(1-y^2\nu^2)+ x^2 y^2 k_0\,,\nonumber\\
 J(x,y)&=&\frac{2 k^2 (1-x^2) (1-y^2) \lambda  \sqrt{\nu}}{(x-y) (1-\nu)^2}
 \,(1+\lambda ^2 -\nu ^2  + 2 (x+y) \lambda  \nu-x y \, k_0,\nonumber\\
 F(x,y)&=& \frac{2 k^2}{(x-y)^2 (1-\nu)^2} (G(x) (1-y^2)\left(\left((1-\nu)^2-\lambda ^2\right)\right.
  (1+\nu )\nonumber\\
      &&\left.+y \lambda (1-\lambda ^2+2 \nu -3 \nu ^2)\right)+ G(y) (2 \lambda ^2+ x \lambda ((1-\nu )^2+\lambda ^2)\nonumber\\
      &&+ x^2\left((1-\nu )^2-\lambda ^2\right)(1+\nu)+x^3 k_1+ x^4 (1-\nu )\, k_0)).\nonumber
   \eeqa
The solution \footnote{We have analytically verified that the solution presented in \cite{Pomeransky:2006bd}
  indeed satisfies the Einstein vacuum equations, $R_{\mu\nu}=0$. Note that this form of the metric is also in \cite{Pomeransky:2006bd} except that we interchange $\phi$ and $\psi$, so that $\phi$ is the
  azimuthal angle of the $S^2$ and $\psi$ parameterizes the circle of the ring} is parameterized by a scale $k$ and two
dimensionless parameters $\lambda$ and $\nu$ which are required to satisfy
$0\leq\nu<1\,$ and $ 2\sqrt{\nu}\leq\lambda<1+\nu$.
The metric has a coordinate singularity at the roots
where $g_{yy}$ diverges. The roots of the equation
$  1+\lambda \, y+\nu \, y^2 = 0$
determine the locations of the inner and outer horizons;
the event horizon is located at
\begin{small}
\be
  y_h=\frac{-\lambda+\sqrt{\lambda^2-4\nu}}{2\nu} \, .
\ee
\end{small}
The properties of these black rings, such as the phase diagrams and limits, were first analyzed in \cite{Elvang:2007hs}. A summary of the results is presented here.
There are three limiting cases of the parameters in this solution: $\nu \to 0$, $\lambda \to 2\nu^{1/2}$ and $\nu \to 1, \lambda \to 2$. On one hand it was found that the latter two limiting cases correspond to extremal solutions.
The extremal black ring with $\lambda = 2 \nu^{1/2}$ is regular and
has zero temperature as expected. Physically it corresponds to the $S^2$ rotating maximally, i.e.~saturating the Kerr bound. There exists zero temperature black rings
for any $S^1$ angular momentum $j_\psi > 3/4$. This is quite similar to the case of supersymmetric black rings \cite{Gauntlett:2004wh,Elvang:2004rt,Elvang:2004ds,Bena:2004de}. Remarkably, it was shown that the entropy of this non-supersymmetric extremal black ring can be reproduced from a microscopic calculation\cite{Reall:2007jv,Emparan:2008qn}.
The limit $\nu \to 1, \lambda \to 2$ appears singular, but this is just a coordinate artifact, and the resulting solution is actually the extremal Myers-Perry black hole with parameters $a_1$, $a_2$ and $\mu^{1/2} = a_1 + a_2$. In this collapse limit the area is discontinuous, just like in the similar collapse limits of supersymmetric black rings \cite{Elvang:2004ds}. So this are endpoints where the ring has collapsed to the zero temperature Myers-Perry black hole.
 On the other hand, when $\nu \to 0$, the solution is the balanced black ring \cite{Emparan:2001wn} with rotation only in the plane. In this case, note that since the balance condition has already been imposed \cite{Pomeransky:2006bd} the unbalanced black ring with angular momentum only on the $S^2$ \cite{Mishima:2005id,Tomizawa:2005wv} cannot be obtained from the
Pomeransky-Sen'kov solution. The more general unbalanced doubly spinning black ring metric \cite{Morisawa:2007di} contains this limit.

Another qualitative feature is the disappearance of the ``fat ring
branch'' as $j_\phi \ge 1/5$, becomes large. Diagonally
opposite 2-spheres of the ring carry $j_\phi$ angular
momentum which creates an attractive spin-spin interaction
\cite{Wald:1972sz}. This is what causes the
diminishing and the disappearance of the fat ring branch as $j_\phi$ increases.

The analysis of \cite{Elvang:2006dd}, in concordance with \cite{Arcioni:2004ww}, suggested that the thin black ring branch
solutions are stable to radial perturbations and the fat rings
unstable. Extrapolating these results, doubly spinning rings with
large enough $S^2$ angular momentum, $j_\phi \ge 1/5$, may be expected to be radially stable.

The physical parameters of the doubly spinning black ring can be written
\begin{subequations}
\label{doublyBR}
\begin{equation}
M =   \frac{3 \pi\, k^2}{G} \frac{\lambda}{1+\nu-\lambda}\qquad ,\,S =
  \frac{8 \pi^2 k^3 \, \lambda (1+\nu+\lambda)}{G(1-\nu)^2(y_h^{-1}-y_h)} \,,
\end{equation}
\begin{equation}
 T_\rom{H} = \frac{(y_h^{-1} - y_h) (1-\nu) \sqrt{\lambda^2 - 4 \nu}}{8\pi
    \, k\, \lambda (1+\nu +\lambda)} \,,
\end{equation}
\begin{small}
\begin{equation}
J_\phi =
  \frac{4 \pi\, k^3}{G}
  \frac{\lambda \, \sqrt{\nu \big((1+\nu)^2 - \lambda^2 \big)}}
         {(1+\nu-\lambda)(1-\nu)^2} \,, \,\Omega_\phi= \frac{\lambda (1+\nu)-(1-\nu)\sqrt{\lambda^2 - 4\nu}}
    {4 k\, \lambda \sqrt{\nu}}
    \sqrt{\frac{1+\nu-\lambda}{1+\nu+\lambda}} \,
\end{equation}
\begin{equation}
  J_\psi =
  \frac{2 \pi\, k^3}{G}
  \frac{\lambda \, (1+\lambda-6 \nu + \nu\, \lambda+\nu^2)\, \sqrt{ (1+\nu)^2 - \lambda^2}}
         {(1+\nu-\lambda)^2(1-\nu)^2} \qquad,\,\Omega_\psi =  \frac{1}{2 k}
  \sqrt{\frac{1+\nu-\lambda}{1+\nu+\lambda}} \,.
\end{equation}
\end{small}
\end{subequations}
Examining the ranges of the angular
momenta one finds that the angular momenta can never be equal, and the ratio
$J_\phi/ J_\psi\le 1/3$.

The ultra-spinning regimes of black rings can be found in \cite{art}.

\vspace{-2mm}\subsubsection*{Black ring in all dimensions}

Heuristically, a black ring can be defined by taking a black string (see the following section), bending and wrapping it into a circle, $S^1$, and spinning it in order to balance its self-gravitational attraction .

The method employed in the construction of the higher dimensional black rings was the matched asymptotic expansion \cite{Gorbonos:2004uc,Gorbonos:2005px}. The general idea was to match the linearized gravity solution for a thin black ring away from the horizon to a near-horizon solution for a bent boosted black string.
An important result of this exercise is that the perturbed event
horizon remains regular.

For the convenience of the reader we collect here the entire thermodynamics :
\begin{subequations}
\label{tbrthermo}
\begin{equation}
\label{tbrthermo1}
M=\frac{\Omega_{n+1}}{8 G}\,R\, r_0^{n}(n+2) \,,\qquad S=\frac{\pi\,\Omega_{n+1}}{2G}  R\,r_0^{n+1}
\sqrt{\frac{n+1}{n}}\,,
\end{equation}
\begin{equation}\label{tbr}
T_H = \frac{n}{4\pi} \sqrt{\frac{n}{n+1}} \frac{1}{r_0}
\,,
\end{equation}
\begin{equation}
\label{tbrthermo2}
J=\frac{\Omega_{n+1}}{8 G}\,R^2\, r_0^{n}\sqrt{n+1} \,,\qquad \Omega_H = \frac{1}{\sqrt{n+1}} \frac{1}{R}\,.
\end{equation}
\end{subequations}
These results are valid up to $O(r_0^2/R^2)$ corrections.

\vspace{-2mm}\subsubsection*{Helical black ring in all dimensions}

Due to its elasticity, the thin black ring can be bent and balanced in an helicoidal shape(a spring ring) as was shown in \cite{Emparan:2009vd}. The horizon being $S^1\times S^{D-3}$, it preserves only two commuting Killing vector fields in agreement with the rigidity theorems of \cite{Hollands:2006rj, Hollands:2010qy}. The physical parameters characterizing the helical black ring are
\begin{subequations}
\label{thbrthermo}
\begin{equation}
\label{thbrthermo1}
M=\frac{\Omega_{n+1}}{8G}(n+2)r_0^n\sqrt{\sum n_a^2 R_a^2}
\;\;,\, S=\frac{\pi\Omega_{n+1}}{2G}\,r_0^{n+1}\sqrt{\sum n_a^2 R_a^2}\sqrt{\frac{n+1}{n}}\,,
\end{equation}
\begin{equation}
T_H=\frac{n}{4\pi} \sqrt{\frac{n}{n+1}} \frac{1}{r_0}
\,,
\end{equation}
\begin{equation}
J_a=\pm\frac{\Omega_{n+1}}{8G}\sqrt{n+1}\,r_0^n \,n_a R_a^2\,,\qquad \Omega_a = \frac{1}{\sqrt{n+1}} \frac{n_a}{\sqrt{\sum n_a^2 R_a^2}}\,,
\end{equation}
\end{subequations}
where at least two strands $n_i>n_j>0$ and all of them integers. The helical black ring with the shortest length is entropically favoured.

\vspace{-2mm}\subsection{Blackfolds}

The horizons topology of blackfolds, p-tuboids or in its more general form as products of odd-spheres, are $\prod_i S^{m_i}  \times S^{D-2-m}$ horizon topologies where $2\leq m_i\leq n$, where $  m_i\in \mathbb{N}_{odd}$ and $m=\sum_i m_i\leq n$. For completeness we present its physical parameters
\begin{subequations}
\begin{equation}
M=\frac{R^m\Omega_m \Omega_{n+1}}{16 \pi G} \, r_0^{n}(n+m+1)
\end{equation}
\begin{equation}
 S=\frac{R^m\Omega_{m}  \Omega_{n+1}}{4G}  r_0^{n+1}
\sqrt{\frac{n+m}{n}} \,,\qquad T = \frac{n}{4\pi} \sqrt{ \frac{n}{n+m}}
\frac{1}{r_0}\,,
\end{equation}
\begin{equation}
J_i=\frac{1}{k+1} \frac{R^{m+1}\Omega_{m}  \Omega_{n+1}}{16\pi G}\,
r_0^{n}\sqrt{m(n+m)}\,,\qquad \Omega_i =  \sqrt{\frac{m}{n+m} }
\frac{1}{R} \, .
\end{equation}
\end{subequations}
where $R$ is the radius of the $m$-sphere. A detailed analysis can be found in \cite{Emparan:2009vd}.

\vspace{-2mm}\section{Multi black holes}\label{sec:multi}

All the examples of higher-dimensional black holes that we have discussed so far present a single event horizon and can, accordingly, be referred to as \emph{uni black holes}. However, unlike its four-dimensional counterpart, higher-dimensional GR also admits black-hole solutions with several, disconnected horizons: the so called \emph{multi black holes}. Examples of multi back holes include a five-dimensional \emph{black saturn} \cite{Elvang:2007rd}, a combination of a black ring with a Myers-Perry black hole at its centre, \emph{di-ring} \cite{Iguchi:2007is,Evslin:2007fv}, a coplanar configuration of two concentric rings or the \emph{bicycling black ring} \cite{Elvang:2007hs}, consisting of two five-dimensional black rings rotating in orthogonal planes. Due to the lengthy expressions for the metrics of this multi black holes we will not include them here.\\
Black holes, and black rings in particular, have been usually found by means of educated guesses. In some cases, however, a systematic procedure to generate black hole solutions can be used. This is the inverse scattering method \cite{Belinsky:1971nt,Belinsky:1979,Belinski:2001ph}.
The underlying idea behind the method is to make use of the complete integrability of the system of non-linear equations that follow from Einstein's equations for solutions with sufficient symmetry. Remarkably, among the solutions with the required degree of symmetry are the rotating black holes in various dimensions. And, perhaps more surprisingly,  the technique can also be used even to generate multi black holes solutions. Note that solutions in curved backgrounds and $D>5$ with asymptotically flat boundary conditions cannot be constructed with the method.

The metric of a $D$-dimensional stationary vacuum space with $D-2$ commuting Killing vector fields can be written in block diagonal form\cite{Emparan:2001wk,Harmark:2004rm}. Furthermore, the two-planes orthogonal to the Killing vector fields are integrable. This means that one can always introduce a coordinate system that is independent of the corresponding $D-2$ coordinates. Taking together all these considerations, the metric can be cast in the following canonical form:
\be
ds^2=G_{ab}(\rho,z) \,dx^adx^b+e^{2\,\nu(\rho,z)}(d\rho^2+dz^2)
\ee
where the conformal factor $e^{2\,\nu(\rho,z)}$ is a function of $\rho, z$ and $G_{ab}(\rho,z)$ is an induced metric in a $D-2$ dimensional hyperplane. Without loss of generality we can take coordinates such that $\det(G_{ab})=-\rho^2$.
The Einstein equations for this kind of metrics decouple and the system is completely integrable. Several strategies were developed to deal with the problem of the appearance of singularities. For singly spinning black holes a uniform rescaling or renormalization was introduced \cite{Mishima:2005id,Tomizawa:2005wv,Koikawa:2005ia,Azuma:2005az,Tomizawa:2006vp}. And, in order to generate healthy solutions with rotations along any number of planes a more general method was proposed in \cite{Pomeransky:2005sj} and applied to generate many multi black holes.
\begin{figure}
\vspace{-4mm}\centerline{\includegraphics[width=11cm,height=4cm]{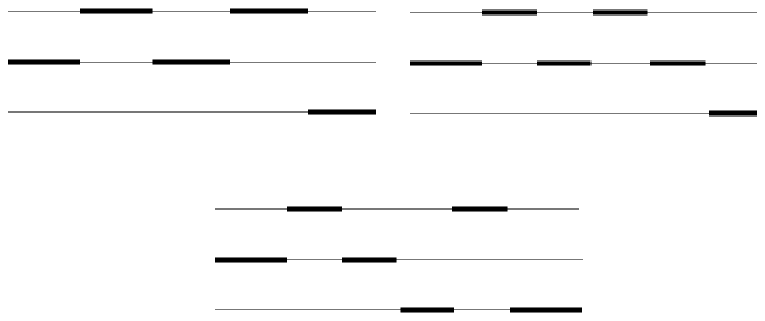}}
\begin{picture}(0,0)(0,0)
\put(74,75){\small Black Saturn}
\put(225,75){\small Black Di-Ring}
\put(140,5){\small Bicycling Black Ring}
\end{picture}
\vspace{-3mm}
\caption{\scriptsize Rod structure of all multi horizon black holes (besides combinations between them) in five dimensions. In the most general solutions the (\textit{upper}) horizon rods have mixed directions $(1,\Omega^{(i)}_{\phi},\Omega^{(i)}_{\psi})$, $i=1,2$ corresponding to each component of the multi black holes solution, while the other (\textit{lower}) rods are oriented purely along $\phi$ and $\psi$.}\label{fig:rods3}\vspace{-3mm}
\end{figure}
A remarkable feature of this type of solutions is that they can be characterized by their \textit{rod structure}, as defined in \cite{Harmark:2004rm} generalizing \cite{Emparan:2001wk}. It involves the specification of the rods and its directions to characterize a solution. A graphical representation of the rod structure that determines each solution uniquely\footnote{The static solutions are characterized uniquely by the rod diagrams. However, a unique characterization of five-dimensional stationary solutions is more subtle \cite{Hollands:2007aj}.} can be found (see Fig. \ref{fig:rods3}).

\vspace{-3mm}\section{The black hole solutions in curved backgrounds}\label{sec:curvedblacksolutions}

This section is devoted to the (A)dS black holes solutions.

\vspace{-2mm}\subsection{(A)dS black holes in all dimensions}

The stationary black hole solution with dS or AdS asymptotics was found in four dimensions \cite{Carter:1968ks}\footnote{The static black hole solution with AdS asymptotics had been found preiously by Kottler \cite{Kottler}.} and thirty years later in five dimensions \cite{Hawking:1998kw}. The extension of this solution, known as the Kerr-de Sitter and Anti-de Sitter metric, to higher dimensions was carried out by Gibbons, Lu, Page and Pope and, in dimension $D$ and Boyer-Lindquist coordinates, is given by \cite{Gibbons:2004uw,Gibbons:2004js}
\beqa\label{KerrAdsSolution}
ds^2 =- W\, (1 -\lambda\,r^2)\,
d\tau^2  +
\frac{2M}{U}\left[W d\tau - \sum_{i=1}^N \frac{a_i\, \mu_i^2\, d\varphi_i}{
\Xi_i}\right]^2 + \frac{U\, dr^2}{V-2M}+r^2 d\alpha^2 \epsilon\nonumber\\
+ \sum_{i=1}^N \frac{r^2 + a_i^2}{\Xi_i}
 [d\mu_i^2 + \mu_i^2 (d\varphi_i +\lambda a_i d\tau)^2] +
\frac{\lambda}{W(1-\lambda r^2)}
\left[\sum_{i=1}^{N+\epsilon} \frac{(r^2 + a_i^2)}{
\Xi_i}\mu_i d\mu_i\right]^2
\eeqa
where $i=1, \ldots , N=[(D-1)/2]$ and $\epsilon=\mod_2 \,\,(D-1)$, so that $D=2 N+1+\epsilon$. There are $N$ the azimuthal angles $\varphi_i$ and
$(N+\epsilon)$ direction cosines $\alpha, \mu_i$ obeying the constrain $\sum_i^{N}\,\mu_i^2 +\epsilon \,\alpha^2=1$. Also the mass parameter $M$
and the rotational parameters $a_i$ are free. Finally, $\lambda=\Lambda/(D-1)$, where $\Lambda$ is the cosmological constant and the functions
$U$, $V$, $W$ and $\Xi_i$ are defined by
\be
W \equiv \sum_{i=1}^{N+\epsilon} \frac{\mu_i^2}{\Xi_i}\,,\qquad U\equiv r^{\epsilon}\sum_{i=1}^{N+\epsilon} \frac{\mu_i^2}{r^2+a_i^2}\prod_{j=1}^{N}(r^2+a_j^2)
\ee
\be
V\equiv r^{\epsilon-2}\, (1-\lambda\, r^2)\, \prod_{i=1}^N (r^2 + a_i^2)\,,\qquad \Xi_i\equiv 1 + \lambda\,a_i^2\,.
\ee
In the limit of vanishing cosmological constant, $\Lambda\rightarrow 0$ (\ref{KerrAdsSolution}) reduces, as expected, to the asymptotically flat MP black hole (\ref{MyersPerrySolution}).

We now discuss some aspects of the (A)dS black hole with more than one angular
momentum\cite{Gibbons:2004uw}. Their mass, angular momenta, area and surface gravity, as computed in \cite{Gibbons:2004ai}, are
\vspace{-4mm}\begin{equation}
M=\frac{m\,{\Omega}_{D-2}}{4\pi\prod_j\Xi_j}
\left(\sum_{i=1}^N\frac1{\Xi_i}-\frac{1-\epsilon}{2}\right),\qquad
J_i=\frac{m\,{\Omega}_{D-2}a_i}{4\pi\Xi_i\prod_j\Xi_j}\,,
\end{equation}
\begin{equation}
{\mathcal A}=\frac{\Omega_{D-2}}{r_+^{1-\epsilon}}
\prod_{i}\frac{r_+^2+a_i^2}{\Xi_i}\,,\,\,\,
\kappa=r_+\lp1+\frac{r_+^2}{L^2}\rp\lp\sum_i\frac1{r_+^2+a_i^2}+\frac\epsilon{2r_+^2}\rp-\frac1{r_+}\,,
\end{equation}
where $N=\left[\frac{D-1}2\right]$ is the maximal number of independent
angular momenta, $a_i$ are the $N$ angular velocity parameters, $m$ is
the mass parameter, $\epsilon=(D-1)\mathrm{mod}\,2$, $\Xi_i=1-{a_i^2}/{L^2}$.
This solution displays a spherical $S^{D-2}$ event horizon situated at $r=r_+$, the highest real solution of $V-2M=0$. Note that this equation is the same as the equation for the horizon of the asymptotically flat Myers-Perry black hole in $D+2$ dimensions, where the additional rotation is $a_{N+1}=L$ and mass parameter is $\mu=2L^2m$. Hence, the root structure and the horizons of the Kerr-AdS$_D$ black hole can be inferred from the MP$_{D+2}$ solution; in particular,
for odd $D$, an horizon always exists provided that any two of the spin parameters vanish, while for even $D$, its existence is guaranteed if any one of the spins vanishes. Therefore, under this assumption, an ultraspinning limit can be achieved for all but two (one) of the $a_i\rightarrow L$ in odd (even) dimensions.

These black holes comply with the ``BPS bound''\cite{Chrusciel:2006zs}
\be
ML\geq \sum_{i=1}^N \left|J_i\right|
\label{posenergy}
\ee

This bound can only be saturated in the ultra-spinning regime, in which one or more spin parameters tend to $L$, but never when all the angular momenta are
non-zero. Indeed, suppose $n$ spin parameters approach the ultraspinning limit. To keep the mass finite, we need to scale the parameters as
\be
\Xi_{\al=1\ldots n}=\xi_\al\nu\,,\qquad m=\mu\nu^{n+1},
\ee
where $\nu\rightarrow0$ in the ultraspinning limit, while keeping $\xi_1,\ldots,\xi_n$ and $\mu$ constant. As we observed previously, this limit is allowed provided any one (two) of the $a_i$ vanish in even (odd) dimensions. Then the root $r_+$ tends to zero, while the mass and angular momenta reach the values (with $\al,\beta=1\ldots n$ running on the spin parameters that tend to $L$ and $I=n+1,\ldots,N$ denoting the others)
\vspace{-2mm}\beqa
M=\frac{\mu\Omega_{D-2}}{4\pi\Pi_\al\xi_\al\Pi_I\Xi_I}\sum_\al\frac1{\xi_\al}\,,\qquad
J_\al=\frac{\mu\Omega_{D-2}}{4\pi\xi_\al\Pi_\beta\xi_\beta\Pi_I\Xi_I}\,,\qquad
J_I=0\,,\vspace{-2mm}
\eeqa
and saturate the BPS bound (\ref{posenergy}). However, these black holes are not extremal, since the surface gravity diverges like $\kappa\rightarrow(2k+\epsilon-2)/2r_+$, where $k$ is the number of vanishing spin parameters. The area vanishes in the limit, decreasing to zero like
\vspace{-3mm}\be
\sa \propto \sm^{\frac{2k+\epsilon-1}{2k+\epsilon-2}}\left(1-\frac{\sj}{\sm}\right)^{\frac{2k+n+\epsilon-1}
{2k+\epsilon-2}}\left(1+\mc{O}(\sm-\sj)\right)\,.\vspace{-2mm}
\ee
The limiting black holes are pancaked out along the planes of rotation (the geometry describes a black membrane with horizon topology $\mathbb{R}^{2n}\times S^{D-2(n+1)}$) and so, it is reasonable to presume that they will develop a Gregory-Laflamme type of instability.
\vspace{-7mm}\subsection{(A)dS black rings in all dimensions}

It is natural to ask whether black rings exist in higher dimensions.
Their existence (or absence) in Anti-de Sitter space is of
special interest for the possible implications in the context of the
AdS/CFT duality. However, in spite of attempts since early on, an exact
solution describing an (Anti-)de Sitter black ring remains elusive.

Nevertheless, there appears no obvious physical reason why these
solutions should not exist. Putting a black ring in Anti-de Sitter space should have the effect of increasing the gravitational centripetal pull on it, but,
at least within some parameter ranges, this can be plausibly balanced by
spinning the black ring faster. On the other hand, if we put the ring in
de Sitter space, the cosmological expansion should act against the
tension, and so the required rotation should be smaller and possibly
reach zero. Thin black rings have been constructed via approximate
methods in every dimension $D\geq 5$ \cite{Emparan:2007wm}. The physical quantities are
\begin{subequations}
\be\label{ringM}
M=\frac{\Omega_{n+1}}{8
G}Lr^{n}_0(n+2)\sr\left(1+\sr^2\right)^{3/2}\,,
\,S=\frac{\pi\Omega_{n+1}}{2G} Lr_0^{n+1}
\sr\sqrt{\frac{n+1+(n+2)\sr^2}{n}}\,,\nonumber\vspace{-2mm}
\ee
\be
T_H=\frac{n^{3/2}\sqrt{1+\sr^2}}{4\pi
r_0\sqrt{n+1+(n+2)\sr^2}}\,,\vspace{-2mm}
\ee
\be\label{ringJ}
J=\frac{r^{n}_0\,L^2}{8
G}\Omega_{n+1}\sr^2 \left[
\left(1+(n+2)\sr^2\right)\left(n+1+(n+2)\sr^2\right)
\right]^{1/2}\,,\vspace{-2mm}
\ee
\begin{equation}
\Omega_H=
\frac{1}{L}
\sqrt{\frac{(1+\sr^{2})(1+(n+2)\sr^2)}{\sr^2(n+1+(n+2)\sr^2)}}\,.
\end{equation}
\end{subequations}
 in principle valid up to corrections of order $r_0/\min{(R,L)}$.

\vspace{-2mm}\section{Transverse asymptotically flat black holes}

We proceed now to recall the metrics for the simplest black strings ($p=1$) and black $p$-branes, characterized by transverse asymptotically flat boundary conditions, and extended horizons with topologies $S^{D-2-p}\times \mathbb{R}^p$.

Black $p$-branes and black strings (1-branes) \cite{Horowitz:1991cd} are $D$-dimensional solutions that arise from the combination of a $D-p$ dimensional Schwarzschild-Tangherlini metric with a flat, Euclidean metric on the remaining $\mathbb{R}^p$. These extended black holes are transverse asymptotically flat (namely, asymptotically flat in only  $D-p$ directions), evade the no-hair theorems, and exhibit horizon topologies $S^{D-2-p}\times \mathbb{R}^p$. Their metric, in $D$ dimensions, is of the form
\be
ds^2 = -V dt^2 +\frac{1}{V} dr^2 + r^2 d\Omega^2_{D-p-2} + dx_i dx^i \,.
\ee
where $V = 1 -(\frac{r_+}{r})^{D-p-3}$ and $i=1,2, \ldots ,p$. The event horizon is situated at $r = r_+ $. The $x^i$ directions correspond to the flat part $\mathbb{R}^p$; alternatively, they can be periodically identified, $x^i=x^i+ 2\pi \,R^i$, yielding the so-called localized black objects in Kaluza-Klein circles (see e.g. the review \cite{Kol:2004ww}).

\vspace{-3mm}\section{The phase diagram}\label{sec:phase}
\begin{figure}
\centerline{\includegraphics[width=6cm,height=3.5cm]{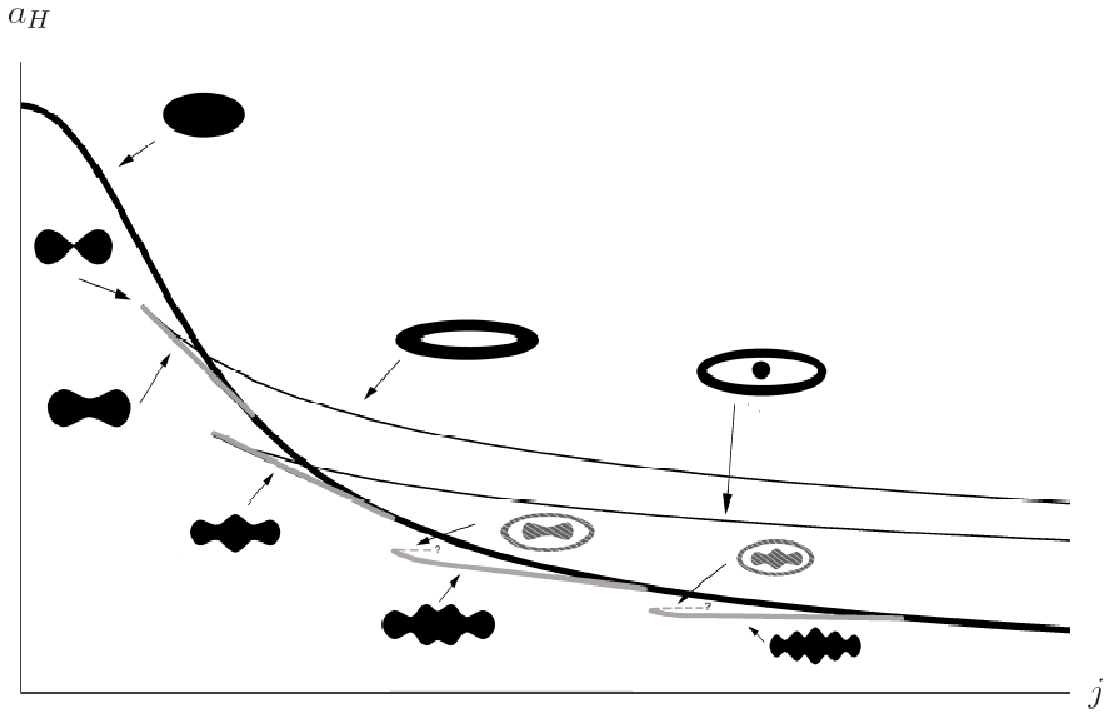}\hspace{0.2cm}
\includegraphics[width=6cm,height=3.5cm]{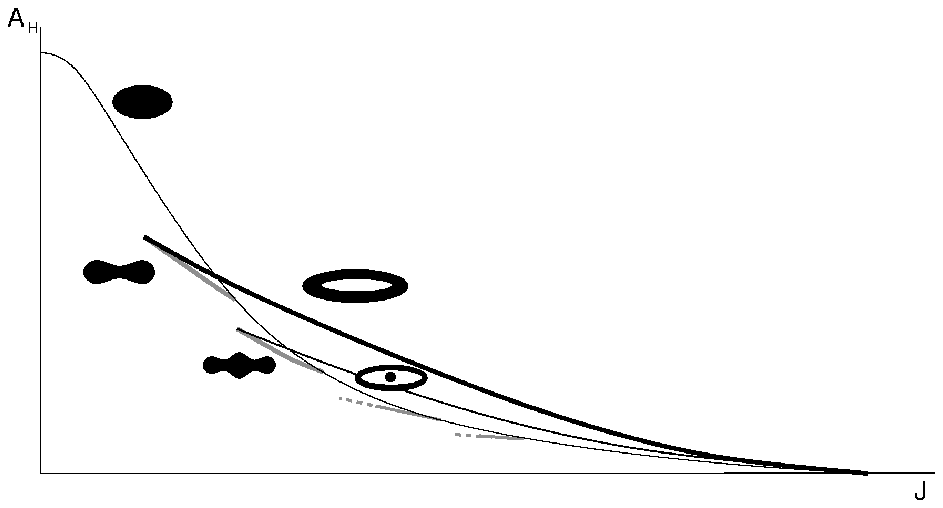}}
\caption{\scriptsize Proposal for the completion of phase curves
in $D\geq 6$. The plot on the left are the patterns for asymptotically flat black holes with a single spin proposed in \cite{Emparan:2007wm}. In AdS, the plot on the right, the pattern is compressed to the range $\sj\leq \sm$ at small $\sm$. We stress that the details of the
connections (\eg first order vs.\ second order transitions) remain
unknown and are arbitrarily drawn.}
\vspace{-5mm}
\label{fig:ads_loM}
\end{figure}

The understanding of the black hole phases in five dimensions has advanced greatly in recent years. As we have seen, besides the well known uni horizon black holes, 
\begin{figure}\vspace{-7mm}
\centerline{\includegraphics[width=7cm,height=4.5cm]{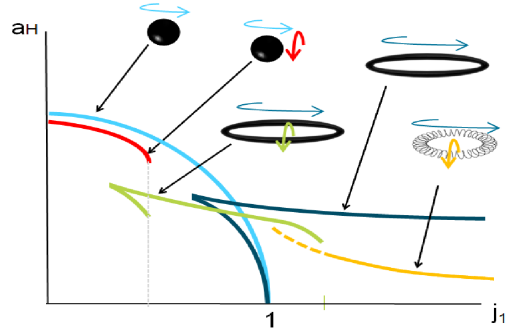}}
\vspace{-2mm}
\caption{\scriptsize Phase diagram of all known uni horizon black holes in five space times dimensions. The doubly spinnign black objects have its second angular momentum equaly fixed $j_2=0.1$ as well as the strands of the helical black ring to maximize the area.}
\vspace{-4mm}
\label{fig:phase5}
\end{figure}
namely the Myers-Perry black hole and the black ring, there also exist multi black hole solutions. In fact, in five dimensions it is possible that essentially all uni or multi horizons black holes with two axial Killing vectors have been found by now (up to iterations between them). All these findings are represented in Fig. \ref{fig:phase5} that include the doubly spinning black objects.
In contrast, the situation in six or more dimensions is much more obscure. Only the MP black hole \cite{Myers:1986un} is explicitly known and the black rings \cite{Emparan:2007wm}, helical black rings and blackfolds \cite{Emparan:2009vd} only perturbatively. For black holes with asymptotically (A)ds boundary conditions exact black holes \cite{Hawking:1998kw,Gibbons:2004uw,Gibbons:2004js} and thin black rings\cite{Caldarelli:2008pz}. These phases, and the proposal of \cite{Emparan:2007wm}, are shown in Fig. \ref{fig:ads_loM}.
The most interesting analysis comes from comparing the different black hole solutions in higher dimensions to elucidate and learn which properties change when tuning the number of dimensions. We present the phase diagrams in the micro-canonical ensemble (area vs. angular momenta with fixed mass) in this section.

\vspace{-4mm}\section{The selection rule}

Black objects in certain regimes have a black membrane phase and behave accordingly. One could then use the inverse logic and build new black holes, by bending these horizons to form compact objects with appropriate boundary conditions, from a black string/brane. This idea was widely employed in \cite{Emparan:2007wm,Emparan:2009vd} for generating \textit{thin} black rings, helical black rings and blackfolds. In the process of constructing the higher dimensional black ring\cite{Emparan:2007wm}, it was found that the absence of naked singularities required a zero-tension condition that corresponds to balancing the string/brane tension against the centrifugal repulsion. In other words, General Relativity encodes(selects) in the equations of motion of black holes the regularity conditions on the geometry.

This condition is in tight correspondence with the conservation of the stress energy tensor. The quasilocal formalism \cite{Brown:1992br,Mann:2005yr,Kraus:1999} gives the appropriate definition for the stress energy tensor in higher dimensions \cite{Astefanesei:2005ad} that, in absence of matter, satisfies a local conservation law
\vspace{-3mm}
\be\label{eq:conservation}
D^a \tau_{ab}=0\vspace{-2mm}
\ee
where the covariant derivative is with respect to the boundary metric $h_{ab}$. The condition (\ref{eq:conservation}) is  then satisfied in the absence of conical singularities \cite{Astefanesei:2009mc}. The conservation and explicit expressions of the stress tensor can be found in \cite{Astefanesei:2009wi}. This extra ingredient is the balance(zero tension) condition encoding the selection rule of GR for regular black hole configurations.\\
As an example we find the balance(zero tension) condition for a D-dimensional black ring with dipole charges, as solution of Einstein-Maxwell-dilaton theory with the dilaton coupling $a=4/N-4n/(n+2)$. At high spin its geometry will be that of a straight black string with boost and charge (parametrized by $\alpha$ and $\gamma$ respectively). Therefore, ($\ref{eq:conservation}$) determines the specific value for the boost parameter required for this agreement between the two geomteries. A straightforward computation fixes 
\vspace{-2mm}\be\label{eq:boost}
\sinh^2 \alpha=(1/n)+N\sinh^2\gamma\vspace{-2mm}
\ee
The charges of the \textsl{thin} D-dipole black ring are the ones of the charged boosted black string with a fixed boost value (\ref{eq:boost}). In $D=5$ these agree with \cite{Emparan:2004wy}.

\vspace{-4mm}\section{Outlook}

We were able to provide a catalogue for current known species of $D$-black holes.
In spite of all this headway, the complete list of all possible topologies that the event horizon of a higher dimensional black hole can display, for each of the three  relevant asymptotic behaviors (Minkowski, AdS and dS), is still unknown. Only few explicit metrics of higher dimensional black holes are known and so, it would be worth completing the task and find the more exotic species. It would be also interesting to investigate the stability and to further explore the selection rule for regular black hole solutions in GR.

\vspace{-4mm}\section*{Acknowledgements}

MJR wants to thank D. Astefanesei, M. Caldarelli, H. Elvang,  R. Emparan, T. Harmark, J. Kunz, R. B. Mann, V. Niarchos, N. Obers, E. Radu, S. Theisen and O. Varela for the valuable discussions.


\vspace{-4mm}\begin{thebibliography}{99}


\bibitem{ArkaniHamed:1998nn}
  N.~Arkani-Hamed, S.~Dimopoulos and G.~R.~Dvali,
  \textit{Phenomenology, astrophysics and cosmology of theories with  sub-millimeter dimensions and TeV scale quantum gravity},
  Phys.\ Rev.\ {\bf D59}, 086004 (1999)
[arXiv:hep-ph/9807344].

\bibitem{Wheeler}
J.~ A.~ Wheeler and K.~ W.~ Ford,
\textit{Geons, Black Holes, and Quantum Foam: A Life in Physics},
W. W. Norton and Company  (2000) and references therein (1968).

\bibitem{Kerr}
Roy P.~ Kerr,
{\it Gravitational Field of a Spinning Mass as an example of Algebraically Special Metrics},
Phys.\ Rev.\ Lett.\ {\bf 11}, 237 (1963).

\bibitem{Mazur:1982db}
  P.~O.~Mazur,
 \textit{Proof Of Uniqueness Of The Kerr-Newman Black Hole Solution},
  J.\ Phys.\ {\bf A15}, 3173 (1982).

\bibitem{BuntingTesis}
G.~L.~ Bunting,
 \textit{ Proof of the uniqueness conjecture for black holes},  PhD thesis,
Umversity of New England, Armidale, NSW, (1983).

\bibitem{Schwarzschild}
Karl Schwarzschild,
{\it On the gravitational field of a mass point according to Einstein's theory},
Sitzungsber.\ Preuss.\ Akad.\ Wiss.\ Berlin (Math.\ Phys.) {\bf 1916} 189 (1916)
[arXiv:physics/9905030].

 \bibitem{Israel}
W.~Israel,
{\it Event Horizon in static Vacuum Space-Time},
Phys.\ Rev.\ {\bf 164}, 1776 (1967).

\bibitem{Carter}
B.~Carter,
{\it Axisymmetric Black Holes Has Only Two Degrees of Freedom},
Phys.\ Rev.\ Lett.\ {\bf 26}, 331 (1971).


\bibitem{Hawking1}
S.~W.~Hawking,
{\it Black Holes in general relativity},
Commun.\ Math.\ Phys.\ {\bf 25}, 152 (1972).


\bibitem{Myers:1986un}
  R.~C.~Myers and M.~J.~Perry,
  \textit{Black Holes In Higher Dimensional Space-Times},
  Annals Phys.\  {\bf 172}, 304 (1986).

\bibitem{Emparan:2001wn}
R.~Emparan and H.~S. Reall,
{\it A rotating black ring in five dimensions},
  Phys.\ Rev.\ Lett.\  {\bf 88}, 101101 (2002)
 [arXiv:hep-th/0110260].


\bibitem{Pomeransky:2006bd}
  A.~A.~Pomeransky and R.~A.~Sen'kov,
  \textit{Black ring with two angular momenta,}
  arXiv:hep-th/0612005.


\bibitem{Elvang:2007rd}
  H.~Elvang and P.~Figueras,
 \textit{ Black Saturn},
  JHEP {\bf 0705}, 050 (2007)
[arXiv:hep-th/0701035].


\bibitem{Iguchi:2007is}
H.~Iguchi and T.~Mishima,
\textit{Black di-ring and infinite nonuniqueness},
  Phys.\ Rev.\ {\bf D75}, 064018 (2007)
[arXiv:hep-th/0701043].


\bibitem{Evslin:2007fv}
J.~Evslin and C.~Krishnan,
\textit{The black di-ring: An inverse scattering construction},
arXiv:0706.1231 [hep-th].

\bibitem{Elvang:2007hs}
  H.~Elvang and M.~J.~Rodriguez,
 \textit{Bicycling Black Rings,}
  JHEP {\bf 0804}, 045 (2008)
  [arXiv:0712.2425 [hep-th]].


\bibitem{Emparan:2007wm}
  R.~Emparan, T.~Harmark, V.~Niarchos, N.~A.~Obers and M.~J.~Rodriguez,
 \textit{The Phase Structure of Higher-Dimensional Black Rings and Black Holes,}
  JHEP {\bf 0710}, 110 (2007)
  [arXiv:0708.2181 [hep-th]].

\bibitem{Emparan:2009cs}
  R.~Emparan, T.~Harmark, V.~Niarchos and N.~A.~Obers,
 \textit{Blackfolds,}
  Phys.\ Rev.\ Lett.\  {\bf 102}, 191301 (2009)
  [arXiv:0902.0427 [hep-th]].


\bibitem{Bunting}
 G.~L.~Bunting and A.~K.~M.~ Masood-ul-Alam,
 {\it Non existence of multiple black holes in asymptotically Euclidean static vacuum space-times},
 Gen.\ Relativ.\ Gravit.\ {\bf 19}, 147 (1987).

\bibitem{Gibbons:2002bh}
  G.~W.~Gibbons, D.~Ida and T.~Shiromizu,
  {\it Uniqueness and non-uniqueness of static vacuum black holes in higher dimensions},
  Prog.\ Theor.\ Phys.\ Suppl.\  {\bf 148}, 284 (2003)
[arXiv:gr-qc/0203004]. Phys.\ Rev.\ Lett.\  {\bf 89}, 041101 (2002)
  [arXiv:hep-th/0206049].

\bibitem{Morisawa:2004tc}
Y.~Morisawa and D.~Ida,
\textit{A boundary value problem for the five-dimensional stationary rotating black holes},
Phys.\ Rev.\ {\bf D69}, 124005 (2004)
[arXiv:gr-qc/0401100].

\bibitem{Hollands:2007aj}
S.~Hollands and S.~Yazadjiev,
\textit{Uniqueness theorem for 5-dimensional black holes with two axial killing fields},
arXiv:0707.2775  [gr-qc].

\bibitem{Galloway:2005mf}
  G.~J.~Galloway and R.~Schoen,
{\it A generalization of Hawking's black hole topology theorem to higher dimensions},
  Commun.\ Math.\ Phys.\  {\bf 266}, 571 (2006)
[arXiv:gr-qc/0509107].

\bibitem{Galloway:2006ws}
  G.~J.~Galloway,
  {\it Rigidity of outer horizons and the topology of black holes},
 arXiv:gr-qc/0608118.


\bibitem{Reall:2002bh}
  H.~S.~Reall,
\textit{Higher dimensional black holes and supersymmetry,}
  Phys.\ Rev.\  D {\bf 68}, 024024 (2003)
  [Erratum-ibid.\  D {\bf 70}, 089902 (2004)]
  [arXiv:hep-th/0211290].

\bibitem{Donaldson}
 S.~ Donaldson and P.~ Kronheimer,
 {\it The Geometry of Four-Manifolds},
  Oxford University Press (1997).
 [arXiv:gr-qc/0605106].

  \bibitem{Helfgott:2005jn}
C.~Helfgott, Y.~ Oz and Y.~ Yanay,
{\it On the topology of black hole horizons in higher dimensions},
JHEP {\bf 0602}, 025 (2006)
[arXiv:hep-th/0509013].

\bibitem{Kleihaus:2009wh}
  B.~Kleihaus, J.~Kunz and E.~Radu,
 \textit{$d\geq 5$ static black holes with $S^2\times S^{d-4}$ event horizon topology,}
  Phys.\ Lett.\  B {\bf 678}, 301 (2009)
  [arXiv:0904.2723 [hep-th]].

\bibitem{Dias:2010eu}
  O.~J.~C.~Dias, P.~Figueras, R.~Monteiro, H.~S.~Reall and J.~E.~Santos,
 \textit{An instability of higher-dimensional rotating black holes,}
  arXiv:1001.4527 [hep-th].

  \bibitem{Kottler}
F.~ Kottler,
{\it The physical basics of Einstein's theory of gravitation},
 Annals.\  Phys.\ {\bf 56},  401 (1918).

\bibitem{Witten:1998zw}
  E.~Witten,
  \textit{Anti-de Sitter space, thermal phase transition, and confinement in  gauge
  theories},
  Adv.\ Theor.\ Math.\ Phys.\  {\bf 2}, 505 (1998)
 [arXiv:hep-th/9803131].

\bibitem{Hawking:1998kw}
  S.~W.~Hawking, C.~J.~Hunter and M.~Taylor-Robinson,
  {\it Rotation and the AdS/CFT correspondence},
  Phys.\ Rev.\ {\bf D59}, 064005 (1999)
 [arXiv:hep-th/9811056].


\bibitem{Gibbons:2004uw}
  G.~W.~Gibbons, H.~Lu, D.~N.~Page and C.~N.~Pope,
  {\it The general Kerr-de Sitter metrics in all dimensions},
  J.\ Geom.\ Phys.\  {\bf 53}, 49 (2005)
 [arXiv:hep-th/0404008].

\bibitem{Gibbons:2004js}
  G.~W.~Gibbons, H.~Lu, D.~N.~Page and C.~N.~Pope,
  {\it Rotating black holes in higher dimensions with a cosmological constant},
  Phys.\ Rev.\ Lett.\  {\bf 93}, 171102 (2004)
 [arXiv:hep-th/0409155].


\bibitem{Caldarelli:2008pz}
  M.~M.~Caldarelli, R.~Emparan and M.~J.~Rodriguez,
\textit{Black Rings in (Anti)-deSitter space,}
  JHEP {\bf 0811}, 011 (2008)
  [arXiv:0806.1954 [hep-th]].


\bibitem{Emparan:2003sy}
R.~Emparan and R.~C. Myers,
{\it Instability of ultra-spinning black holes},
  JHEP {\bf 09}, 025 (2003)
[arXiv:hep-th/0308056].


\bibitem{art}
  D.~Astefanesei, M.~J.~Rodriguez and S.~Theisen,
\textit{Thermodynamic instability of doubly spinning black objects,} \textit{to appear}.


\bibitem{Dias:2009iu}
  O.~J.~C.~Dias, P.~Figueras, R.~Monteiro, J.~E.~Santos and R.~Emparan,
\textit{Instability and new phases of higher-dimensional rotating black holes,}
  arXiv:0907.2248 [hep-th].


\bibitem{Emparan:2006mm}
R.~Emparan and H.~S. Reall,
\textit{Black rings},
 Class.\ Quant.\ Grav.\ {\bf 23}, R169 (2006)
[arXiv:hep-th/0608012].

\bibitem{Belinsky:1971nt}
V.~A. Belinsky and V.~E. Zakharov,
\textit{Integration of the Einstein
equations by the inverse scattering problem technique and the calculation of the exact soliton solutions},
  Sov.\ Phys.\ JETP {\bf 48}, 985 (1978)
[Zh.\ Eksp.\ Teor.\ Fiz.\  {\bf 75} (1978) 1953].

\bibitem{Belinsky:1979}
V.~A. Belinsky and V.~E. Zakharov,
\textit{Stationary gravitational solitons with axial symmetry},
Sov.\ Phys.\ JETP {\bf 50}, 1 (1979)
   [Zh.\ Eksp.\ Teor.\ Fiz.\  {\bf 77} (1979) 3].

\bibitem{Belinski:2001ph}
V.~Belinski and E.~Verdaguer,
\textit{Gravitational solitons},
Cambridge University Press {\bf 258} (2001).

\bibitem{Gauntlett:2004wh}
  J.~P.~Gauntlett and J.~B.~Gutowski,
  \textit{Concentric black rings},
  Phys.\ Rev.\ {\bf D71}, 025013 (2005)
[arXiv:hep-th/0408010].


\bibitem{Elvang:2004rt}
  H.~Elvang, R.~Emparan, D.~Mateos and H.~S.~Reall,
  \textit{A supersymmetric black ring},
  Phys.\ Rev.\ Lett.\  {\bf 93}, 211302 (2004)
[arXiv:hep-th/0407065].


\bibitem{Elvang:2004ds}
  H.~Elvang, R.~Emparan, D.~Mateos and H.~S.~Reall,
  \textit{Supersymmetric black rings and three-charge supertubes},
  Phys.\ Rev.\ {\bf D71}, 024033 (2005)
 [arXiv:hep-th/0408120].
\bibitem{Bena:2004de}
  I.~Bena and N.~P.~Warner,
\textit{One ring to rule them all ... and in the darkness bind them?},
  Adv.\ Theor.\ Math.\ Phys.\  {\bf 9}, 667 (2005)
 [arXiv:hep-th/0408106].

\bibitem{Reall:2007jv}
  H.~S.~Reall,
  \textit{Counting the microstates of a vacuum black ring},
arXiv:0712.3226 [hep-th].

\bibitem{Emparan:2008qn}
  R.~Emparan,
\textit{Exact Microscopic Entropy of Non-Supersymmetric Extremal Black Rings,}
  Class.\ Quant.\ Grav.\  {\bf 25}, 175005 (2008)
  [arXiv:0803.1801 [hep-th]].


\bibitem{Mishima:2005id}
  T.~Mishima and H.~Iguchi,
  \textit{New axisymmetric stationary solutions of five-dimensional vacuum  Einstein
  equations with asymptotic flatness},
  Phys.\ Rev.\ {\bf D73}, 044030 (2006)
[arXiv:hep-th/0504018].

\bibitem{Tomizawa:2005wv}
  S.~Tomizawa, Y.~Morisawa and Y.~Yasui,
\textit{Vacuum solutions of five dimensional Einstein equations generated by
  inverse scattering method},
  Phys.\ Rev.\ {\bf D73}, 064009 (2006)
[arXiv:hep-th/0512252].

\bibitem{Morisawa:2007di}
  Y.~Morisawa, S.~Tomizawa and Y.~Yasui,
  \textit{Boundary Value Problem for Black Rings},
arXiv:0710.4600 [hep-th].


\bibitem{Wald:1972sz}
R.~Wald,
\textit{Gravitational spin interaction},
Phys.\ Rev.\ {\bf D6}, 406 (1972).


\bibitem{Elvang:2006dd}
  H.~Elvang, R.~Emparan and A.~Virmani,
 \textit{ Dynamics and stability of black rings},
  JHEP {\bf 0612}, 074 (2006)
  [arXiv:hep-th/0608076].


\bibitem{Arcioni:2004ww}
  G.~Arcioni and E.~Lozano-Tellechea,
  \textit{Stability and critical phenomena of black holes and black rings},
  Phys.\ Rev.\ {\bf D72}, 104021 (2005)
[arXiv:hep-th/0412118].


\bibitem{Gorbonos:2004uc}
D.~Gorbonos and B.~Kol,
\textit{A dialogue of multipoles: Matched asymptotic expansion for caged black holes},
 JHEP {\bf 06}, 053 (2004)
[arXiv:hep-th/0406002].

\bibitem{Gorbonos:2005px}
D.~Gorbonos and B.~Kol,
\textit{Matched asymptotic expansion for caged black holes: Regularization of the post-Newtonian order},
 Class.\ Quant.\ Grav.\  {\bf 22}, 3935 (2005)
[arXiv:hep-th/0505009].

\bibitem{Emparan:2009vd}
  R.~Emparan, T.~Harmark, V.~Niarchos and N.~A.~Obers,
  ``New Horizons for Black Holes and Branes,''
  arXiv:0912.2352 [hep-th].

\bibitem{Hollands:2006rj}
  S.~Hollands, A.~Ishibashi and R.~M.~Wald,
  {\it A higher dimensional stationary rotating black hole must be axisymmetric},
  Commun.\ Math.\ Phys.\  {\bf 271}, 699 (2007)

\bibitem{Hollands:2010qy}
  S.~Hollands, J.~Holland and A.~Ishibashi,
  ``Further restrictions on the topology of stationary black holes in five
  dimensions,''
  arXiv:1002.0490 [gr-qc].

\bibitem{Emparan:2001wk}
  R.~Emparan and H.~S.~Reall,
  \textit{Generalized Weyl solutions},
  Phys.\ Rev.\ {\bf D65}, 084025 (2002)
 [arXiv:hep-th/0110258].

\bibitem{Harmark:2004rm}
  T.~Harmark,
  \textit{Stationary and axisymmetric solutions of higher-dimensional general relativity},
  Phys.\ Rev.\ {\bf D70}, 124002 (2004)
[arXiv:hep-th/0408141].


\bibitem{Koikawa:2005ia}
  T.~Koikawa,
\textit{Infinite number of soliton solutions to 5-dimensional vacuum Einstein equation},
  Prog.\ Theor.\ Phys.\  {\bf 114}, 793 (2005)
 [arXiv:hep-th/0501248].


\bibitem{Azuma:2005az}
  T.~Azuma and T.~Koikawa,
  \textit{Infinite number of stationary soliton solutions to five-dimensional vacuum Einstein equation},
  Prog.\ Theor.\ Phys.\  {\bf 116}, 319 (2006)
[arXiv:hep-th/0512350].
\bibitem{Tomizawa:2006vp}
  S.~Tomizawa and M.~Nozawa,
  \textit{Vaccum solutions of five-dimensional Einstein equations generated by
  inverse scattering method. II: Production of black ring solution},
  Phys.\ Rev.\ {\bf D73}, 124034 (2006)
 [arXiv:hep-th/0604067].

\bibitem{Pomeransky:2005sj}
A.~A. Pomeransky, 
\textit{Complete integrability of higher-dimensional
Einstein equations with additional symmetry, and rotating black holes},
Phys.\ Rev.\ {\bf D73}, 044004 (2006)
[arXiv: hep-th/0507250].


\bibitem{Carter:1968ks}
  B.~Carter,
  \textit{Hamilton-Jacobi and Schrodinger separable solutions of Einstein's equations},
  Commun.\ Math.\ Phys.\  {\bf 10}, 280 (1968).



\bibitem{Chrusciel:2006zs}
  P.~T.~Chrusciel, D.~Maerten and P.~Tod,
 \textit{ Rigid upper bounds for the angular momentum and centre of mass of
  non-singular asymptotically anti-de Sitter space-times},
  JHEP {\bf 0611}, 084 (2006)
  [arXiv:gr-qc/0606064].

\bibitem{Gibbons:2004ai}
  G.~W.~Gibbons, M.~J.~Perry and C.~N.~Pope,
  \textit{The first law of thermodynamics for Kerr-anti-de Sitter black holes},
  Class.\ Quant.\ Grav.\  {\bf 22}, 1503 (2005)
  [arXiv:hep-th/0408217].

\bibitem{Horowitz:1991cd}
  G.~T.~Horowitz and A.~Strominger,
  \textit{Black strings and P-branes},
  Nucl.\ Phys.\ {\bf B360}, 197 (1991).

\bibitem{Kol:2004ww}
B.~Kol,
\textit{The phase transition between caged black holes and black strings: A review},
 Phys.\ Rept.\ {\bf 422}, 119 (2006)
[arXiv:hep-th/0411240].
\bibitem{Brown:1992br}
  J.~D.~Brown and J.~W.~.~York,
\textit{Quasilocal energy and conserved charges derived from the gravitational
  action,}
  Phys.\ Rev.\  D {\bf 47}, 1407 (1993)
  [arXiv:gr-qc/9209012].

\bibitem{Mann:2005yr}
  R.~B.~Mann and D.~Marolf,
\textit{Holographic renormalization of asymptotically flat spacetimes,}
  Class.\ Quant.\ Grav.\  {\bf 23}, 2927 (2006)
  [arXiv:hep-th/0511096].

   \bibitem{Kraus:1999}
   P.~Kraus, F.~Larsen and R.~Siebelink,
\textit{The gravitational action in asymptotically AdS and flat spacetimes,}
  Nucl.\ Phys.\  B {\bf 563}, 259 (1999)
  [arXiv:hep-th/9906127].


\bibitem{Astefanesei:2005ad}
  D.~Astefanesei and E.~Radu,
\textit{Quasilocal formalism and black ring thermodynamics,}
  Phys.\ Rev.\  D {\bf 73}, 044014 (2006)
  [arXiv:hep-th/0509144].

\bibitem{Astefanesei:2009mc}
  D.~Astefanesei, M.~J.~Rodriguez and S.~Theisen,
\textit{Quasilocal equilibrium condition for black ring,}
  JHEP {\bf 0912}, 040 (2009)
  [arXiv:0909.0008 [hep-th]].

\bibitem{Astefanesei:2009wi}
  D.~Astefanesei, R.~B.~Mann, M.~J.~Rodriguez and C.~Stelea,
\textit{Quasilocal formalism and thermodynamics of asymptotically flat black
  objects,}
  arXiv:0909.3852 [hep-th].

\bibitem{Emparan:2004wy}
  R.~Emparan,
\textit{Rotating circular strings, and infinite non-uniqueness of black rings,}
  JHEP {\bf 0403}, 064 (2004)
  [arXiv:hep-th/0402149].

\end{thebibliography}
\end{document}